\documentclass[sigconf, review，authordraft]{acmart}
\usepackage{algorithm}
\usepackage{algorithmic}
\usepackage{placeins}
\usepackage[english]{babel}
\usepackage{hyphenat}
\usepackage{graphicx}

\usepackage{algorithm} 
\usepackage{algorithmic} 
\usepackage{amsmath} 
\usepackage{subcaption}
\usepackage{caption}
\usepackage{multirow}
\usepackage{tabularx}
\usepackage{longtable}

\AtBeginDocument{%
  }

\setcopyright{acmlicensed}
\copyrightyear{2018}
\acmYear{2018}
\acmDOI{XXXXXXX.XXXXXXX}

\acmConference[Conference acronym 'XX]{Make sure to enter the correct
  conference title from your rights confirmation emai}{June 03--05,
  2018}{Woodstock, NY}
\acmISBN{978-1-4503-XXXX-X/18/06}

\begin{document}

\title{Graph-based Diffusion Model for Collaborative Filtering}

\author{Xuan Zhang}

\email{xuan-zha23@mails.tsinghua.edu.cn}
\affiliation{%
  \institution{the Department of Automation, Tsinghua
University, Beijing National Research Center for
Information Science and Technology}
\country{China}
}
\author{Xiang Deng}
\email{dx23@mails.tsinghua.edu.cn}
\affiliation{%
  \institution{the Department of Automation, Tsinghua
University}
\country{China}
}

\author{Hongxing Yuan}
\email{yuanhx24@mails.tsinghua.edu.cn}
\affiliation{%
  \institution{the Department of Automation, Tsinghua
University, Beijing National Research Center for
Information Science and Technology}
\country{China}
}

\author{Chunyu Wei}
\authornotemark[1]
\email{weichunyu@ruc.edu.cn}

\affiliation{%
  \institution{the School of Information, Renmin University of China}
  \country{China}
}
\author{Yushun Fan}

\authornotemark[1]
\email{fanyus@tsinghua.edu.cn}
\affiliation{%
  \institution{the Department of Automation, Tsinghua
University, Beijing National Research Center for
Information Science and Technology}
\country{China}
}

\renewcommand{\shortauthors}{X.~Zhang et al.}

\begin{abstract}
Recently, diffusion-based recommendation methods have achieved impressive results. However, existing approaches predominantly treat each user's historical interactions as independent training samples, overlooking the potential of higher-order collaborative signals between users and items. Such signals, which encapsulate richer and more nuanced relationships, can be naturally captured using graph-based data structures. 
To address this limitation, we extend diffusion-based recommendation methods to the graph domain by directly modeling user-item bipartite graphs with diffusion models. This enables better modeling of the higher-order connectivity inherent in complex interaction dynamics. However, this extension introduces two primary challenges: (1) Noise Heterogeneity, where interactions are influenced by various forms of continuous and discrete noise, and (2) Relation Explosion, referring to the high computational costs of processing large-scale graphs. 
To tackle these challenges, we propose a Graph-based Diffusion Model for Collaborative Filtering (GDMCF). To address noise heterogeneity, we introduce a multi-level noise corruption mechanism that integrates both continuous and discrete noise, effectively simulating real-world interaction complexities. To mitigate relation explosion, we design a user-active guided diffusion process that selectively focuses on the most meaningful edges and active users, reducing inference costs while preserving the graph's topological integrity. 
Extensive experiments on three benchmark datasets demonstrate that GDMCF consistently outperforms state-of-the-art methods, highlighting its effectiveness in capturing higher-order collaborative signals and improving recommendation performance. 
\end{abstract}

\ccsdesc[500]{Information systems~Recommender systems}
\ccsdesc[300]{Information systems~Retrieval models and ranking}

\keywords{Generative Recommender Model, Diffusion Model, Graph Neural Networks}


\maketitle
\section{Introduction}

\label{sec:introduction}
Recommender systems enhance user experiences by suggesting products and content that align with individual preferences, effectively mitigating information overload across various online domains, including e-commerce platforms  \cite{greenstein2018personal}, advertising suggestions, and news websites  \cite{wu2022feedrec}. In contrast to traditional discriminative model-based recommender systems  \cite{chen2020revisiting, zhou2018deep, he2017neural}, generative recommender models based on Generative Adversarial Networks (GANs)  \cite{goodfellow2020generative} or Variational Auto-Encoders (VAEs) \cite{kingma2013auto} hypothesis that user-item interactions are correlated with various latent factors, such as user preferences and product popularity. Due to their superior capacity to model the joint distribution of these complex latent factors, generative recommender systems have demonstrated significant advancements 
 \cite{bharadhwaj2018recgan,chen2019generative,nguyen2023poisoning}.
 
\begin{figure}
	\centering
    \includegraphics[width=1\linewidth]{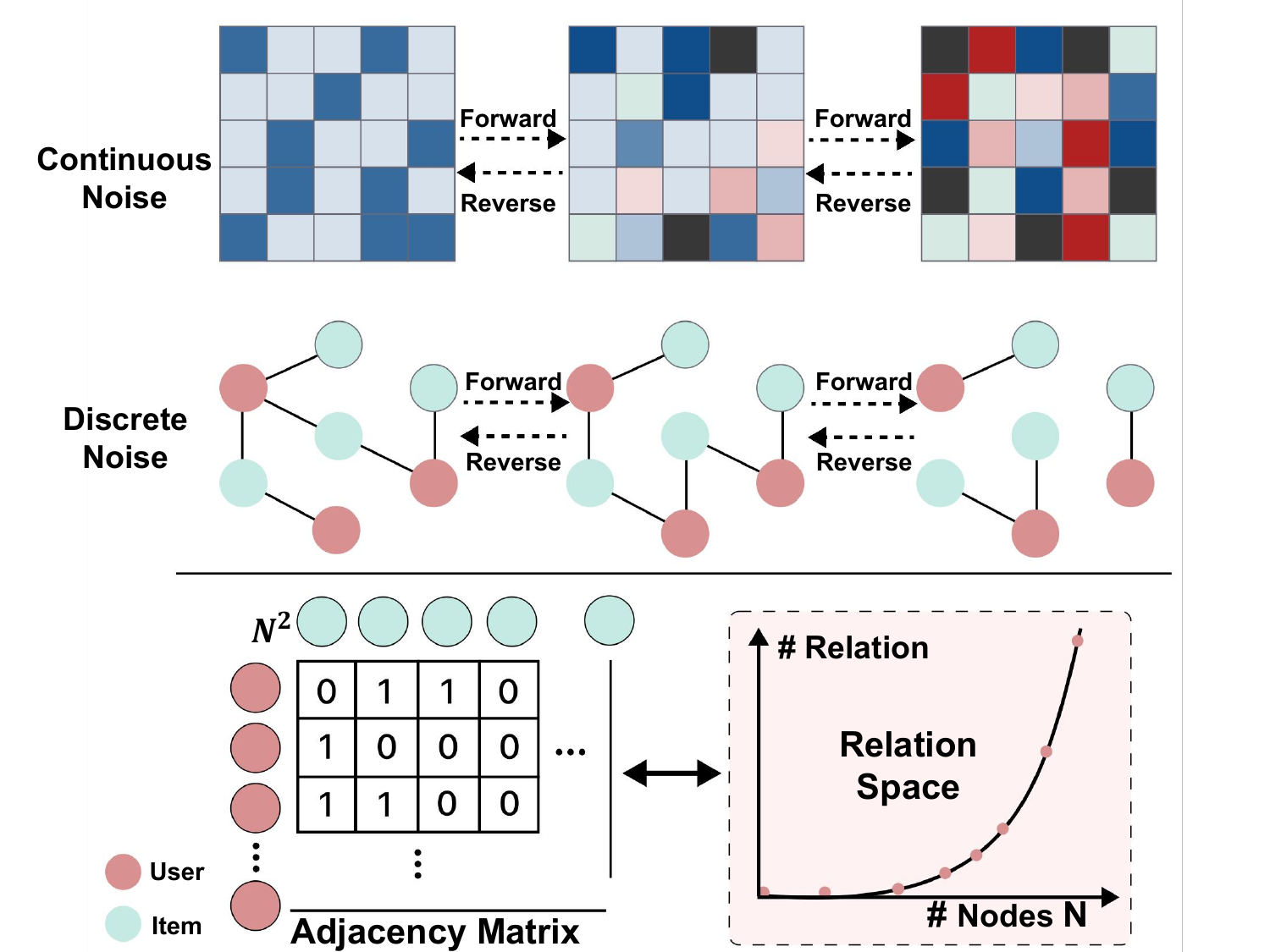}
	\caption{Recommender systems involve various forms of
heterogeneous noise (continuous noise and discrete
noise). As the number of users and items increases,  the complexity of their interactions can grow explosively.}
	\label{fig1}
\end{figure} 
\setlength{\textfloatsep}{10pt}

Recently, the notable successes of diffusion models (DMs)~\cite{ho2020denoising} in high-quality image generation tasks have inspired researchers to explore their application in recommender systems \cite{wu2019neural,li2023diffurec,liu2023diffusion}. Compared to traditional generative methods, DMs provide a highly stable training process \cite{choi2022perception} and can be interpreted through score matching \cite{hyvarinen2005estimation, vincent2011connection, kadkhodaie2021stochastic} and Langevin dynamics \cite{neal2011mcmc,welling2011bayesian}. Additionally, they can be understood from the perspective of diffusion probabilistic models \cite{sohl2015deep,ho2020denoising}, which define a forward diffusion process to add noise to data and a reverse process to recover it.

However, most existing diffusion-based recommendation methods treat a period of interaction sequence 
 \cite{qin2023diffusion,ma2024plug,yang2024survey,lin2024survey} or historical interaction sequence \cite{wang2023diffusion} as a single training sample. These methods overlook the higher-order collaborative signals between users and items in complex recommender systems. For example, if two users have interacted with many of the same items, it suggests they share similar preferences. Additionally, longer interaction chains, such as $i_1 \leftarrow u_2 \leftarrow i_3  \leftarrow u_4$ implies that $u_4$ is likely to show interest in $i_1$, as a similar user $u_2$ has already interacted with $i_1$ \cite{wang2019neural}. Therefore, we argue that it is crucial to explicitly consider these higher-order connections in diffusion-based models focused on interaction generation. Graphs, as a data structure, are inherently suited to modeling relationships between entities. Numerous studies have incorporated graphs into recommender systems, employing techniques such as random walks \cite{fouss2007random, lee2011random, zhang2021graph, choi2022s} or multi-layer graph convolutions 
  \cite{he2020lightgcn,ying2018graph,wang2019knowledge} to effectively capture higher-order collaborative signals. Consequently, we believe that introducing graph-based topological constraints in diffusion-based recommendation methods can better model the higher-order connectivity in complex interaction generation. To achieve this goal, we need to address two notable challenges, as illustrated in Fig. \ref{fig1}.

 \begin{itemize}
    \item \textbf{Noise Heterogeneity}. User-item interactions are inherently complex, shaped by both discrete and continuous factors. Discrete feedback, such as clicks or purchases, reflects binary decisions driven by historical interactions, while continuous variables, like interaction duration, capture nuanced preferences. These combined factors form a joint distribution that requires heterogeneous noise modeling. Consequently, most existing diffusion models \cite{wang2023diffusion,li2023diffurec} relying on a single noise type may be inadequate for the complex dynamics of recommendation scenarios.
     \item \textbf{Relation Explosion}. Traditional graph-based diffusion models require the computation of a latent vector or probability for each node pair in the graph, resulting in a computational complexity of $O(N^2)$. In recommender systems, where the number of users and items is typically very large, this $O(N^2)$ complexity poses a significant challenge to the application of traditional graph-based diffusion models.
 \end{itemize}

To this end, we propose a Graph-based Diffusion Model for Collaborative Filtering (GDMCF) to address the aforementioned challenges synergistically. 

To address the issue of \textbf{Noise Heterogeneity}, we propose a multi-level corruption to capture various forms of noise present in complex recommender systems during the \textit{forward process}. 
For user feature-level perturbations, we introduce continuous noise to corrupt their feature vectors, simulating varying intensities of perturbations in real-world scenarios.
For structure-level perturbations, we introduce discrete noise to corrupt interactions between users and items. 
To ensure consistency in denoising, we first employ an alignment module within the denoising network to integrate the corrupted features and topological structure into a unified corrupted graph. Subsequently, through multiple layers of graph convolution, we iteratively aggregate the higher-order collaborative signals, thereby enhancing graph denoising capabilities. This approach effectively models heterogeneous noise in complex recommender systems by merging both levels into a unified space, thus facilitating joint denoising.

To address the issue of \textbf{Relation Explosion}, we propose a user-active guided generation strategy to enhance the \textit{reverse process} of the diffusion model. On one hand, we preserve the original graph's structural information, specifically the user node degree distribution, allowing us to edit edges during the forward process without compromising the integrity of the original graph structure. On the other hand, during the reverse process, we identify active users based on their degree distribution and retain only the corresponding edges, discarding those associated with inactive users. This ensures that computational resources are prioritized for the most important edges and users, making the model well-suited for large-scale recommendation scenarios. The contributions of this paper are summarized as follows:
\begin{itemize}
     \item We propose a novel Graph-based Diffusion Model for Collaborative Filtering (GDMCF) that applies multi-level corruption to capture heterogeneous noise in real-world scenarios while accounting for the higher-order connectivity in complex interaction generation. 
     
     \item We incorporate a user-active guided generation strategy to more effectively alleviate the heavy computational burden associated with iterative refinement in the reverse process of the diffusion model.
     \item Extensive experiments demonstrate that our method outperforms state-of-the-art baselines across three benchmark datasets. The comprehensive analysis and experimental results confirm the computational efficiency of our approach.
 \end{itemize}

\section{Related Work}
\label{sec:related_work}

\subsection{Generative recommendation}
Discriminative recommendation models predict the probability of interactions between users and items \cite{liu2021interest,wei2022causal}. While existing discriminative models are relatively easy to train, generative recommendation models better capture the underlying data distribution and complex, non-linear relationships \cite{liu2019deep,yuan2019simple}. Most generative models can be broadly categorized into two main types: VAE-based approaches 
  \cite{luo2020deep,shenbin2020recvae}, proposed by 
 \cite{kingma2013auto}, which have demonstrated effectiveness in capturing the latent structure of user interactions by learning an encoder for posterior estimation 
  \cite{nema2021disentangling} and a decoder for predicting interaction probabilities across all items \cite{wang2022causal}. GAN-based models, exemplified by the work of 
 \cite{goodfellow2014generative}, further improve recommendation quality by addressing data sparsity and cold-start issues through adversarial training that refines the recommendation distribution \cite{wu2019pd,yang2018knowledge,gao2019drcgr,he2018adversarial}. Additionally, generative retrieval models have been explored in sequential recommendation tasks, demonstrating their potential to handle sequential dependencies and improve recommendation accuracy \cite{DBLP:conf/nips/RajputMSKVHHT0S23, DBLP:conf/kdd/DeldjooHMKSRVSK24, DBLP:conf/recsys/PenhaVPNB24}.


Recently, diffusion model-based recommender systems have emerged as a superior approach for their stability and high-quality generation. By iteratively reducing noise, they offer improved robustness and flexibility over traditional generative models, resulting in more accurate and diverse recommendations \cite{li2023diffurec,jiang2024diffkg,ma2024plug}.

\subsection{Diffusion Models}
A central challenge in generative modeling is balancing flexibility and computational feasibility\cite{DBLP:journals/corr/abs-2110-10863}. The core concept of diffusion models addresses this by systematically perturbing the data distribution through a forward diffusion process, followed by learning the reverse process to restore the data distribution 
 \cite{ho2020denoising}. This approach results in a generative model that is both highly flexible and computationally efficient 
 \cite{croitoru2023diffusion,yang2023diffusion}. Existing diffusion models can be categorized into two types: conditional generation 
 \cite{tashiro2021csdi,zhang2023shiftddpms,rombach2022high,EDGE} and unconditional generation \cite{ho2022video, austin2021structured}.

Although diffusion models are closely related to other research areas, such as computer vision \cite{songdenoising}, NLP 
 \cite{li2022diffusion}, and signal processing 
 \cite{adib2023synthetic}, their progress in the field of personalized recommendation has been relatively slow. DiffRec \cite{wang2023diffusion} employs a forward noise addition and reverse denoising process on user interaction histories to generate recommendations. It treats the user interaction sequence as a single sample, neglecting the higher-order collaborative signals between users and items. DiffRec adds Gaussian noise to the user-item sequence step by step and utilizes Multilayer Perceptrons (MLPs) for denoising, but the simple MLP structure often fails to capture crucial higher-order signals for accurate recommendations. Our GDMCF addresses this by implementing a graph-based diffusion model that captures higher-order signals in interactions. Recent studies have increasingly focused on the integration of higher-order information within diffusion processes to enhance recommendation performance. GiffCF \cite{DBLP:conf/sigir/ZhuWZX24} leverages heat equations and filtering mechanisms on an item-item graph, effectively capturing item relationships through a diffusion framework. In contrast, GDMCF explicitly models the diffusion process on the user-item bipartite graph, enabling it to directly capture user-item interactions and their propagation. Meanwhile, CF-Diff \cite{DBLP:conf/sigir/HouPS24} introduces a collaborative filtering method based on diffusion models, however, it does not explicitly incorporate diffusion within a graph structure. Instead, CF-Diff relies on rule-based encoding, such as counting incoming links from $(h-1)$-hop neighbors, to extract higher-order information. This approach is inherently static and non-learnable, limiting its flexibility and adaptability to complex recommendation scenarios. Additionally, some studies have examined diffusion processes within social networks \cite{wu2020diffnet++,chen2023group}, primarily investigating how social connections influence user preferences under single-type noise \cite{rafailidis2017recommendation}. These methods differ from GDMCF by focusing on social influence instead of direct user-item diffusion.


\subsection{Graph-based recommendation}

Graph Neural Networks (GNNs) have been essential in leveraging graph-structured data for recommendation tasks 
 \cite{fan2019graph,wu2022graph, chang2021sequential}. The field has evolved from simple random walks \cite{choi2022s} to Graph Convolutional Networks (GCNs) \cite{wang2019knowledge} and attention mechanisms-based methods \cite{chen2019semi}. Random walks 
 \cite{baluja2008video} captured basic relationships but struggled with complex dependencies. GCNs \cite{he2020lightgcn, liu2021interest,zheng2021dgcn} improved this by aggregating neighborhood information across layers, effectively capturing higher-order collaborative signals. Attention mechanism-based methods further enhance this by dynamically weighting nodes and edges \cite{yue2022af}, refining the capture of critical higher-order signals. However, most diffusion-based recommendation methods 
 \cite{wang2023diffusion,li2023diffurec, ma2024plug} often overlook these higher-order collaborative signals in complex recommender systems. GDMCF introduces graph topology constraints in diffusion-based recommendation methods to better model the higher-order connectivity.
\section{Preliminary}
\label{sec:definition}
\subsection{Diffusion Model}
In this section, we introduce the core concepts of diffusion models (DMs), which comprise both forward and reverse processes.

\paragraph{Forward Process}
\sloppy
Given an input data sample $\mathbf{y}^0$, the forward process is defined by $q(\mathbf{y}^t|\mathbf{y}^{t-1})$, which incrementally corrupts $\mathbf{y}^0$ over $T$ steps by adding noise points $(\mathbf{z}^1,\dots,\mathbf{z}^T)$. This process exhibits a Markov structure, where $q(\mathbf{y}^1,\dots,\mathbf{y}^T|\mathbf{y}^0) = q(\mathbf{y}^1|\mathbf{y}^0)\prod_{t=2}^{T} q(\mathbf{y}^t|\mathbf{y}^{t-1})$. As $T \rightarrow \infty$,  $q(\mathbf{y}^T)$ approaches a convergent distribution. 
\paragraph{Reverse Process}
The denoising model $\phi_{\theta}$ is trained to learn the reverse distribution $p_{\theta}(\mathbf{y}^{t-1}|\mathbf{y}^t)$ from $\mathbf{y}^t$ to $\mathbf{y}^{t-1}$. The model follows the joint distribution $p_\theta(\mathbf{y}^{0:T}) = p(\mathbf{y}^T)\prod_{t=1}^{T} p_\theta (\mathbf{y}^{t-1} \mid \mathbf{y}^t)$. $p(\mathbf{y}^T)$ is the convergent distribution in $q$. To generate new samples, noise is sampled from a prior distribution and then progressively inverted using a denoising model. Formally, both Gaussian noise and Bernoulli noise conform to this distribution.

Generally, an effective denoising model must satisfy three key conditions: 1) The distribution $q(\mathbf{y}^t|\mathbf{y}^0)$ should have a closed-form formula, allowing for parallel training across different time steps. 2) The posterior $p_{\theta}(\mathbf{y}^{t-1}|\mathbf{y}^t) = \int q(\mathbf{y^{t-1}}|\mathbf{y}^t, \mathbf{y}^0) dp_{\theta}(\mathbf{y}^0)$ should be expressed in closed form, enabling $\mathbf{y}^0$ to be the target for the denoising model. 3) The limit distribution $q_{\infty} = \lim_{T \rightarrow \infty} q(\mathbf{y}^T|\mathbf{y}^0)$ should be independent of $\mathbf{y}^0$ to be the prior distribution for inference.
\begin{figure*}
	\centering
	\includegraphics[width=1\linewidth]{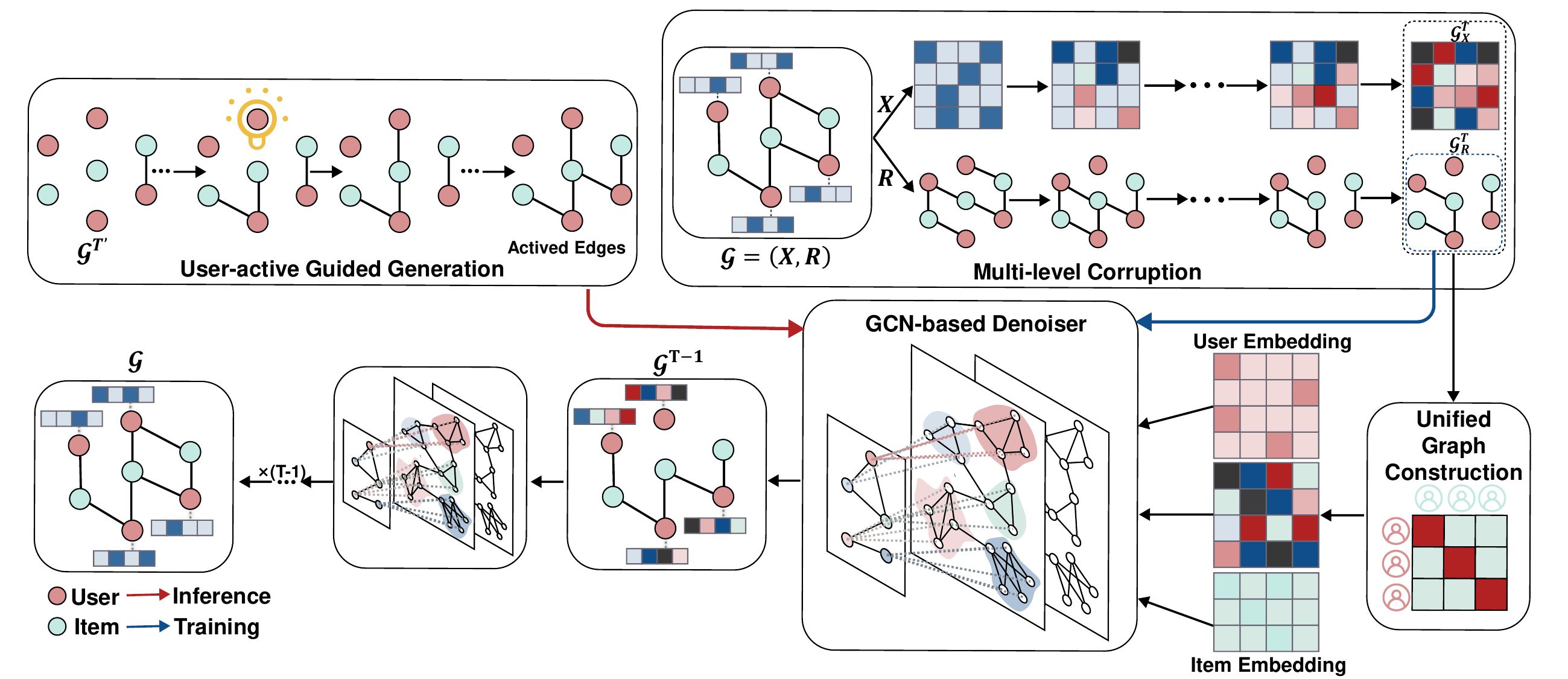}
	\caption{The Framework of GDMCF.}
	\label{fig:GDMCF}
\end{figure*}

\subsection{Recap GCN}
\sloppy
Let $|\mathbf{\mathcal{U}}| = M$ and $|\mathbf{\mathcal{I}}| = N$ represent the sizes of the user set and the item set in recommender systems, respectively. The interaction matrix $\mathbf{R} \in \mathbb{R}^{M \times N}$ denotes the interactions between users and items, where the entries of the matrix are defined as follows:
\begin{equation}
r_{ui}=
\begin{cases}
1& \text{if user $u$ interacted with item $i$},\\
0& \text{otherwise},
\end{cases}
\end{equation}
we define the set of all users and items as $\mathcal{V} = \mathcal{U} \cup \mathcal{I}$, and the edge set as $\mathcal{E} = \left\{(u, i) \mid r_{ui}=1, u \in \mathcal{U}, i \in \mathcal{I}\right\}$, where each edge $(u, i)$ represents an interaction between user $u$ and item $i$. We construct a bipartite graph $\mathcal{G} = (\mathbf{X}, \mathbf{R})$, where $\mathbf{X} = [x_{u1}, \dots, x_{uM}] \in \mathbb{R}^{d_x \times M}$ represents the feature matrix of the users in $\mathcal{U}$, with $d_x$ denoting the feature dimension. The adjacency matrix $\mathbf{A}_{\mathcal{G}}$ for the graph $\mathcal{G}$ is defined as:

\begin{equation}
	\mathbf{A}_{\mathcal{G}}=\left(\begin{array}{cc}
		0 &
\mathbf{R} \\
		
\mathbf{R}^{T} & 0
	\end{array}\right).
\end{equation}
The degree matrix $\mathbf{D}_{\mathcal{G}} \in \mathbb{N}^{(M+N)\times(M+N)}$ of $\mathcal{G}$ is a diagonal matrix, the diagonal element $d_{ii}$ represents the number of non-zero entries in the i-th row vector of $\mathbf{A}_{\mathcal{G}}$. In Graph Convolutional Networks (GCNs) with $L$ layers, the representation of an ego node is updated by aggregating information from its neighboring nodes:

\begin{equation}
	\mathbf{Z}^{(l)}=F(\mathbf{Z}^{(l-1)}, \mathcal{G}),
\end{equation}
here, $\mathbf{Z}^{(l)}$ represents the user or item representations at the $l$-th layer, and $\mathbf{Z}^{(l-1)}$ is the representations of the previous layer. The function $F$ denotes the aggregation function used to aggregate information from neighboring nodes. 

\section{Methodology}



The overall framework of our proposed method is illustrated in Fig. \ref{fig:GDMCF}. We introduce independent corruptions at both the structural and feature levels of the bipartite graph (Sec. \ref{sectionA}), aiming to simulate the inherent noise in real-world recommender systems. A graph-based denoising network then aligns the user and item representations from corruptions, utilizing a GCN-based architecture to reconstruct the corrupted information (Sec. \ref{sectionB}). To address the challenge of relation explosion in large-scale recommender systems, we propose a user-active guided generation strategy (Sec. \ref{sectionC}), which selectively retains edges for activated users, effectively reducing computational complexity and enhancing inference efficiency. Further details on the training optimization and inference procedures are provided in Sec. \ref{sectionD} and Sec. \ref{sectionE}.


\subsection{Multi-level Corruption}
\label{sectionA}

In real-world recommendation scenarios, noise is inherently heterogeneous. 
Existing diffusion-based methods typically rely on single-noise modeling, which makes it challenging to address the complexity and diversity inherent in recommendation tasks. To overcome this limitation, we propose a novel approach that incorporates two distinct types of noise: \textit{Discrete Corruption} and \textit{Continuous Corruption}, integrated into the forward diffusion process. These corruptions are independently applied at the structural and feature levels of the bipartite graph, thereby generating two complementary views that model topological and numerical corruptions, respectively. This intentional design enables our method to effectively capture and accommodate the diverse forms of heterogeneous noise present in real-world recommender systems, enhancing the robustness of the model and improving overall performance.

\subsubsection{Discrete Corruption}
\label{section4.1.1}
To model topological corruption within the bipartite graph, we utilize a probabilistic transition matrix based on the marginal distributions of edge types. Specifically, we define a transition matrix $\boldsymbol{Q}^t$, where $\left[\boldsymbol{Q}^t\right]_{\mathbf{e},\mathbf{e^{\prime}}} = q\left(e^t = \mathbf{e^{\prime}} \mid e^{t-1} = \mathbf{e}\right)$ represents the probability of an edge transitioning from state $\mathbf{e}$ at time step $t-1$ to state $\mathbf{e^{\prime}}$ at time step $t$.
Here, $\mathbf{e}$ and $\mathbf{e^{\prime}}$ denote the two possible edge states (e.g., $[1, 0]$ and $[0, 1]$), corresponding to the absence and presence of an edge, respectively. This transition matrix is parameterized to reflect the marginal distributions of edge types, thereby ensuring that the corrupted graph maintains statistical consistency with the original bipartite graph structure. Let \(\boldsymbol{m} \in \mathbb{R}^{c \times c}\) represents the marginal distribution matrix of edge types in the original bipartite graph $\mathcal{G}^0_R$. The transition matrix is defined as:

\begin{equation}
\boldsymbol{Q}^t=\alpha^t_R \boldsymbol{I}+\beta^t_R \mathbf{1}_c \boldsymbol{m},
\end{equation}
where $\alpha^t_R$ and $\beta^t_R$ are step-dependent weights, $\boldsymbol{I}$ is the identity matrix, $\mathbf{1}_c$ is an all-ones matrix of size $c \times c$, and $c$ represents the number of edge types. 
We transform the original binary interaction matrix $\mathbf{R}$ of a bipartite graph $\mathcal{G} = (\mathbf{X}, \mathbf{R})$ into one-hot encoded vectors. This results in a new matrix $\overrightarrow{\mathbf{R}} \in \mathbb{R}^{M \times N \times 2}$, where $\overrightarrow{\mathbf{R}}_{u,i} = [1, 0]$ indicates the absence of an edge, i.e., $r_{ui} = 0$, and $\overrightarrow{\mathbf{R}}_{u,i} = [0, 1]$ corresponds to the presence of an edge, i.e., $r_{ui} = 1$. Given a bipartite graph $\mathcal{G}^0_R = \mathcal{G}$, the transition to $\mathcal{G}^t_R$ at time $t$ is defined as:

\begin{equation}
q\left(\overrightarrow{\mathbf{R}}^t \mid \overrightarrow{\mathbf{R}}^{t-1}\right)=\overrightarrow{\mathbf{R}}^{t-1} \boldsymbol{Q}^t,
\end{equation}

\begin{equation}
\mathcal{G}^t_R =(\boldsymbol{X}^{0}, \mathbf{R}^{t}).
\label{eq:82}
\end{equation}
Using the reparameterization trick and the multiplicativity of $\boldsymbol{Q}^t$, $\overrightarrow{\mathbf{R}}^t$ is derived from $\overrightarrow{\mathbf{R}}^0$ as:

\begin{equation}
q\left(\overrightarrow{\mathbf{R}}^t \mid \overrightarrow{\mathbf{R}}^0\right)=\overrightarrow{\mathbf{R}}^0 \widebar{\boldsymbol{Q}}^t,
\label{eq:8}
\end{equation}
since \((\boldsymbol{1}_c\boldsymbol{m})^2 = \boldsymbol{1}_c\boldsymbol{m}\), we have $\widebar{\boldsymbol{Q}}^t = \boldsymbol{Q}^1 \cdots \boldsymbol{Q}^t = \widebar{\alpha}^t_R \boldsymbol{I} + \widebar{\beta}^t_R \mathbf{1}_c \boldsymbol{m}$, where $ \widebar{\alpha}_R^t =\prod_{t'=1}^t \alpha^{t'}_R \in (0,1)$ and $\widebar{\beta}_R^t = 1 - \widebar{\alpha}_R^t$, when $T \rightarrow \infty$, $\widebar{\boldsymbol{Q}}^T$ approaches $\mathbf{1}_c\mathbf{m}$.
By defining transition matrices that align with the real data probabilities, we can effectively introduce discrete corruption to the structure of the bipartite graph. The corrupted bipartite graph after $T$ steps is denoted as $\mathcal{G}^T_R$.

\subsubsection{Continuous Corruption}
In addition to modeling topological graph noise, we also model continuous noise at the user feature level, which accounts for numerical corruption in recommendations.
Specifically, given the initial feature $\mathbf{X}^0\in\mathcal{G}$, the transition to $\mathbf{X}^t$ and $\mathcal{G}^t_X$ at time $t \in ({1, \dots, T})$ are represented as follows:
 
\begin{align}
q(\boldsymbol{X}^t \mid \boldsymbol{X}^{t-1}) &= \mathcal{N}\left(\boldsymbol{X}^t ; \sqrt{1-\beta^t_X} \boldsymbol{X}^{t-1}, \beta^t_X \boldsymbol{I}\right),
\label{eq:19}
\end{align}
\begin{align}
\mathcal{G}^t_X = (\boldsymbol{X}^t, \boldsymbol{R}^0) ,
\label{eq:234}
\end{align}
here, $\beta^t_X \in(0,1)$ controls the scale of Gaussian noise added at step $t$. Using the additivity of independent Gaussian noise and the reparameterization trick, $\boldsymbol{X}^t$ is directly derived from $\boldsymbol{X}^0$:
\begin{equation}
q\left(\boldsymbol{X}^t \mid \boldsymbol{X}^0\right)=\mathcal{N}\left(\boldsymbol{X}^t ; \sqrt{\widebar{\alpha}^t_X} \boldsymbol{X}^0,\left(1-\widebar{\alpha}^t_X\right) \boldsymbol{I}\right),
\label{eq:5}
\end{equation}
where $\alpha^t_X=1-\beta^t_X, \widebar{\alpha}^t_X=\prod_{t^{\prime}=1}^t \alpha^{t^{\prime}}_X$, through reparameterization, we obtain $\boldsymbol{X}^t=\sqrt{\widebar{\alpha}^t_X} \boldsymbol{X}^0+\sqrt{1-\widebar{\alpha}^t_X} \boldsymbol{\epsilon} \text { with } \boldsymbol{\epsilon} \sim \mathcal{N}(\mathbf{0}, \boldsymbol{I}) \text {. }$ If $T \rightarrow \infty$, the $\boldsymbol{X}^t$ approaches a standard Gaussian distribution. The corrupted bipartite graph at $T$ step is denoted
as $\mathcal{G}^T_X$. 
By simultaneously modeling topological and numerical corruption as complementary perspectives, our method is able to effectively capture the complex dynamics of user-item interactions in recommender systems.

\subsection{Graph-based Denoising Network}
\label{sectionB}


%
Previous studies \cite{lin2024survey, wang2023diffusion} primarily utilized small multi-layer perceptron (MLP) networks for denoising tasks, often overlooking the higher-order structural information embedded in the data. This limitation restricted their ability to model complex dependencies effectively. To overcome this challenge, we introduce a novel graph-based denoising framework that fully exploits the higher-order information within bipartite graphs. Our approach begins with a unified graph construction module that integrates corrupted node features with topological structure to generate a unified corrupted graph representation. An iterative Graph Convolutional Network (GCN)-based denoiser is then employed to progressively remove noise and refine both structural and feature representations. By combining higher-order information with iterative denoising, our framework captures intricate node relationships and delivers superior denoising performance.

\paragraph{Unified Graph Construction} In recommender systems, it is intuitive that the topological structure of a bipartite graph is heavily influenced by the feature attributes of users and items. Conversely, the features of users and items also reveal certain properties of the bipartite graph topology. This bidirectional relationship highlights the necessity of ensuring consistency between feature-level and topological-level attributes within the bipartite graph. Moreover, it is equally critical to maintain diversity in the corruption process to enable robust learning. To address this, we leverage recent advancements in contrastive learning \cite{chen2020simple, gutmann2010noise} to design a unified graph construction module. Specifically, we transform \( \overrightarrow{\mathbf{R}}^T \in \mathbb{R}^{M \times N \times 2} \) of $\mathcal{G}^T_R$ into \( \mathbf{R}^T \in \mathbb{R}^{M \times N} \) by sampling to ensure a diverse graph structure. Subsequently, we process two distinct corrupted views of the graph: \( \mathcal{G}^T_R \), which incorporates discrete corruption, and \( \mathcal{G}^T_X \), which applies continuous corruption through linear projection layers. This results in two separate sets of user features, \( \boldsymbol{X}^{\prime} \) and \( \boldsymbol{X}^{\prime \prime} \), each capturing unique characteristics associated with their respective types of corruption. We perform representation learning on $\boldsymbol{X}^{\prime}$ and $\boldsymbol{X}^{\prime \prime}$ to aligning user features. The same user in both matrices forms positive pairs $({(x_{u}^{\prime} \in \boldsymbol{X}^{\prime}, x_{u}^{\prime \prime} \in \boldsymbol{X}^{\prime \prime}) \mid u \in \mathcal{U}})$, while different users form negative pairs $({(x_{u}^{\prime} \in \boldsymbol{X}^{\prime}, x_{v}^{\prime \prime} \in \boldsymbol{X}^{\prime \prime}) \mid u,v \in \mathcal{U}})$. The formula in Eq. \ref{eq:1} forces positive user pairs' representations to be close.

\begin{equation}
\mathcal{L}_{ugc} = \sum_{u \in \mathcal{U}} -\log \frac{\exp \left(s\left(x_{u}^{\prime}, x_{u}^{\prime \prime}\right) / \tau\right)}{\sum_{v \in \mathcal{U}} \exp \left(s\left(x_{u}^{\prime}, x_{v}^{\prime \prime}\right) / \tau\right)},
\label{eq:1}
\end{equation}
here, \(s(\cdot)\) denotes the similarity between two representations, and \(\tau\) is the temperature hyperparameter of the Softmax function. The unified user feature matrix \( \widebar{\mathbf{X}}^T \) is constructed by concatenating the aligned features \( \boldsymbol{X}^{\prime} \) and \( \boldsymbol{X}^{\prime \prime} \) along with a corresponding learnable user embedding. The unified corrupted bipartite graph \( \mathcal{G}^T \) is constructed by \( \widebar{\mathbf{X}}^T \) and the interaction matrix \( \mathbf{R}^T \) from \( \mathcal{G}^T_R \). This graph and a learnable item embedding $\mathbf{I}$ are subsequently input into a GCN-based denoiser, which applies a reverse process to iteratively generate a clean bipartite graph.





\paragraph{GCN-based Denoiser}
Modeling the intricate dependencies in a graph often requires leveraging higher-order information. To address this, we propose a GCN-based denoiser designed to model the distribution \( p_\theta \) and effectively perform denoising, enabling the recovery of complex relationships between users and items. Specifically, the GCN denoiser takes \( \mathbf{Z}_T^{(0)} = \text{concat}(\widebar{\mathbf{X}}^T, \mathbf{I})\) as the input node representations and \( \mathbf{A}^T \) as the graph structure, enabling the model to aggregate information across nodes. Through iterative message passing, the GCN refines the representations of each node layer by layer. The representation of each node at layer \( l \), denoted \( \mathbf{Z}_t^{(l)} \), is computed using the propagation rule in Eq.~\ref{eq:16}.

\begin{equation}
\begin{aligned}
\mathbf{Z}^{(l)}_t&=F(\mathbf{Z}^{(l-1)}, \mathcal{G}^t)\\
&= (\mathbf{D}^t)^{-\frac{1}{2}} \mathbf{A}^t (\mathbf{D}^t)^{-\frac{1}{2}} \mathbf{Z}^{(l-1)}_t,
\end{aligned}
\label{eq:16}
\end{equation}
here, \( \mathbf{A}^T \) and \( \mathbf{D}^T \) denote the adjacency matrix and the degree matrix derived from \( \mathbf{R}^T \) of \( \mathcal{G}^T_R \). After applying multiple layers of graph convolution, we obtain the refined node representations $\mathbf{Z}_t$ for the bipartite graph. We divide $\mathbf{Z}_t$ into $P_t$ and $Q_t$, the similarity between $P_t$ and $Q_t$ is measured using the cosine similarity metric for loss calculation, which is defined as:  
\begin{equation}
\begin{aligned}
\text{CosineSim}(P_t, Q_t) = \frac{P_t \cdot Q_t}{\|P_t\| \|Q_t\|}.
\end{aligned}
\label{eq:166}
\end{equation}
%

%
\paragraph{Reverse process}
In summary, the denoiser $\phi_\theta$ models the reverse process, which progressively predicts the target graph $\widehat{\mathcal{G}}$ from the input graph $\mathcal{G}^T$ at each step. The joint probability  $p_\theta\left(\mathcal{G}^{t-1} \mid \mathcal{G}^t\right)$ can be decomposed into a product over users and edges as follows:
\begin{equation}
\begin{aligned}
p_\theta\left(\mathcal{G}^{t-1} \mid \mathcal{G}^t\right) &= p_\theta\left(\widebar{\mathbf{X}}^{t-1} \mid \widebar{\mathbf{X}}^t\right) 
p_\theta\left(\mathbf{R}^{t-1} \mid \mathbf{R}^t \right) \\
&= \prod_{\widebar{x} \in \mathcal{U}} p_\theta\left(\widebar{x}^{t-1} \mid \widebar{x}^t\right) 
\prod_{r \in \mathcal{E}} p_\theta\left(r^{t-1} \mid r^t\right).
\end{aligned}
\label{eq15}
\end{equation}
\begin{algorithm} \renewcommand{\algorithmicrequire}{\textbf{Input:}} \renewcommand{\algorithmicensure}{\textbf{Output:}} \caption{GDMCF Training} \label{algorithmic3} \begin{algorithmic}[1] \REQUIRE User-item interaction bipartate graph $\mathcal{G} = (\mathbf{X}, \mathbf{R})$ and randomly initialized $\theta$, diffusion steps $T$ \REPEAT \STATE Sample $t \sim \mathcal{U}(1, \ldots, T)$, $\boldsymbol{\epsilon} \sim \mathcal{N}(\mathbf{0}, \mathbf{I})$, and $\mathbf{Q}^t$ \STATE Sample $\mathcal{G}^{t}_R$ and $\mathcal{G}^{t}_X$ given $\mathcal{G}$, $t$, $\boldsymbol{\epsilon}$, and $\mathbf{Q}^t$ via Eq. \ref{eq:82} and Eq. \ref{eq:234} \STATE Compute $\widehat{\mathcal{G}}$ through $\phi_\theta$ via Eq.~\ref{eq:16} and Eq.~\ref{eq:166}
\STATE Compute $\mathcal{L}_{\text{ugc}}$ using Eq. \ref{eq:1} and $\mathcal{L}_{\text{diff}}$ using Eq. \ref{eq:56} \STATE Take gradient descent on $\nabla_\theta (\lambda_1\mathcal{L}_{\text{ugc}} + \mathcal{L}_{\text{diff}})$ to optimize $\theta$ \UNTIL{convergence} \ENSURE Optimized $\theta$ \end{algorithmic} \end{algorithm}


\begin{algorithm} 
 \renewcommand{\algorithmicrequire}{\textbf{Input:}}
 \renewcommand{\algorithmicensure}{\textbf{Output:}}
 \caption{GDMCF Inference} 
 \label{alg2} 
 \begin{algorithmic}[1]
     \REQUIRE User-item interaction bipartite graph $\mathcal{G}=(\mathbf{X}, \mathbf{R})$, parameters $\theta$, diffusion steps $T$
\STATE Initialize an empty graph $\mathcal{G}^{T^\prime}$ with $\mathbf{R}^{T^\prime}$
 \FOR {$t = T$ to $1$}
        \STATE Sample $\mathcal{G}^{t}_R$ and $\mathcal{G}^{t}_X$ given $\mathcal{G}$, $t$, $\boldsymbol{\epsilon}$, and $\mathbf{Q}^t$ via Eq. \ref{eq:82} and Eq. \ref{eq:234} 
        \STATE $\mathbf{R}^{(t-1)^\prime} = \text{User-active} (\mathcal{G}^{t^\prime})$ according to Sec. \ref{sectionC}
        \STATE Compute $\widehat{\mathcal{G}}^{t-1}$ through $\phi_\theta$ with $\mathbf{R}^{(t-1)^\prime}$, $\mathcal{G}^{t}_R$ and $\mathcal{G}^{t}_X$
     \ENDFOR
     \ENSURE Generated $\widehat{\mathcal{G}}$


 \end{algorithmic} 
\end{algorithm}
The underlying assumption is that each user's features and corresponding edges evolve independently based on the corrupted graph from the previous time step. This distribution is used to sample $\mathcal{G}^{t-1}$ at each step until $\mathcal{G}$ is generated. 
\begin{figure}
	\centering
	\includegraphics[width=0.95\linewidth]{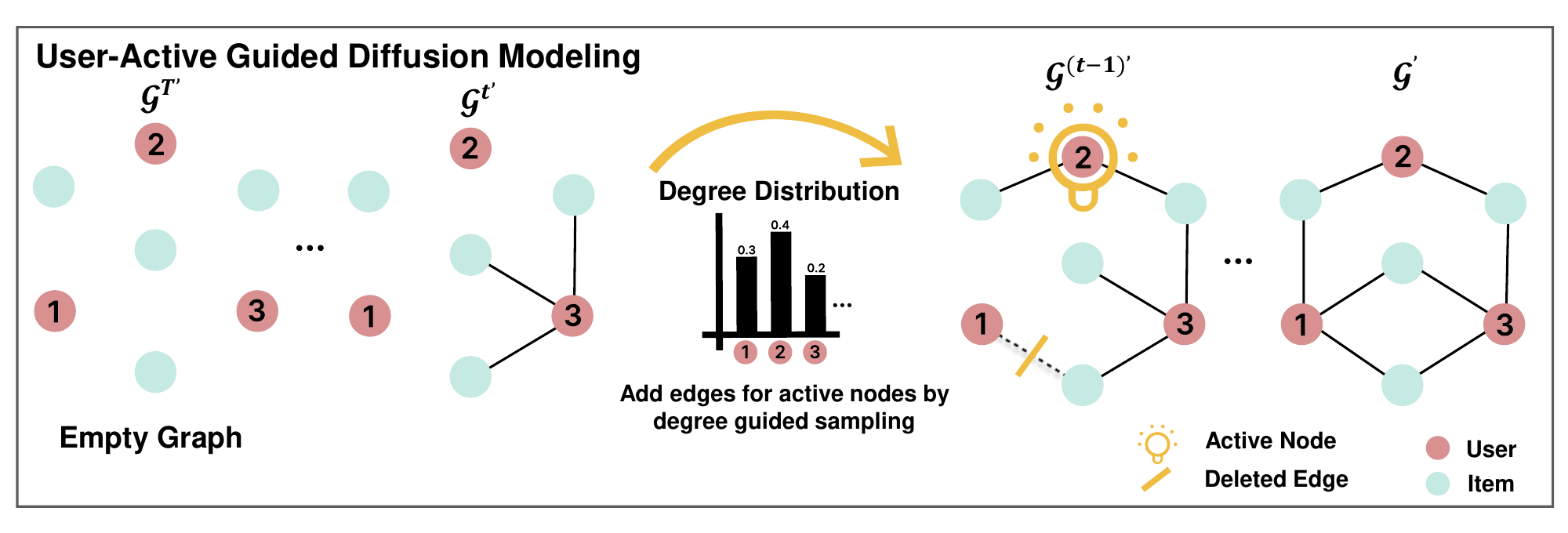}
	\caption{User-Active Guided Diffusion Modeling. In the inference process, GDMCF iteratively adds edges from $\mathcal{G}^{T^{\prime}}$ based on the original graph's degree distribution. The probabilities (0.3, 0.4, 0.2) determine if a user node is activated at step $t$, with only activated nodes' edges retained (e.g., node 2 has a 0.4 probability of retaining its edges).
 }
	\label{fig:diffusion1}
\end{figure}

\subsection{User-active Guided Generation}
\label{sectionC}

While the training process accurately captures higher-order information, the inference process typically requires multiple iterative steps. The aforementioned method requires calculating almost all edges in the interaction matrix, making it impractical for large-scale recommendation scenarios. Inspired by the principles of progressive denoising, where the process begins with a fully noisy initial state and iteratively refines toward the target distribution, we propose a user-active guided generation strategy to improve inference efficiency. Specifically, instead of using the complete bipartite graph throughout the reverse process, we start with an initially empty bipartite graph $\mathcal{G}^{T ^ \prime}$ and continuously add edges under the guidance of user activity. At each step $t$, edges are added using the transition matrix $\boldsymbol{Q}^t$, as defined in Sec. \ref{section4.1.1}, while only the edges connected to activated users are retained. To ensure that the inferred graph gradually approximates the structure of the original graph as the process converges, the active users at step $t$ are determined based on the degree distribution of the original graph $\mathcal{G}$. As illustrated in Fig.~\ref{fig:diffusion1}, the activation probability of user $u$ is determined by the ratio of its degree to the maximum degree of the original graph. The interaction matrix $\mathbf{R}^{t^\prime}$, after edge addition, replaces $\mathbf{R}^{t}$ (as described in Sec.~\ref{sectionB}) as the input to the GCN-based denoiser for prediction. With such a strategy, edges are gradually added, preserving the original graph's statistical properties while significantly reducing computational complexity. Consequently, the reverse process in Eq.~\ref{eq15} is redefined as follows:
\begin{equation}
\begin{aligned}
p_{\theta}\left(\mathcal{G}^{t-1} \mid \mathcal{G}^t\right) &= p_{\theta}(\widebar{\mathbf{X}}^{t-1} \mid \mathbf{X}^{t}) \cdot p_{\theta}(\mathbf{R}^{t-1} \mid \mathbf{R}^{t}, \mathbf{s}^t) \\
&= \prod_{\widebar{x} \in \mathcal{U}} p_\theta\left(\widebar{x}^{t-1} \mid \widebar{x}^{t}\right) \cdot \prod_{r \in \mathcal{E}} p_\theta\left(r^{t-1} \mid r^{t}, \mathbf{s}^t\right),
\end{aligned}
\label{eq:144}
\end{equation}
where $\mathbf{s}^t$ is a binary vector indicating whether a user is activated from $t$ to $t-1$ based on the degree distribution of $\mathcal{G}$. Specifically, the \(u\)-th element of \(\mathbf{s}^t\), denoted as \(s_u^t\), is determined by the following sampling procedure:

\begin{equation}
\begin{aligned}
s_u^t = 
\begin{cases} 
1 & \text{with probability } \frac{d_u^t}{D} \\ 
0 & \text{with probability } 1 - \frac{d_u^t}{D},
\end{cases}
\end{aligned}
\label{eq:148}
\end{equation}
where \(d_u^t\) represents the degree of user \(u\) in the graph \(\mathcal{G}\), and \(D\) denotes the maximum degree of  \(\mathcal{G}\).

\subsection{Optimization}
\label{sectionD}
We optimize the denoiser model $\phi_\theta$ by minimizing the diffusion loss $\mathcal{L}_{\text{diff}} = \sum_{t=1}^T \mathcal{L}_t$, and the calculation of $\mathcal{L}_t$ is as follows:

\begin{equation}
\mathcal{L}_{t}=\mathbb{E}_{\boldsymbol{q}\left(\mathcal{G}^{t} \mid \mathcal{G}\right)}(\|\widehat{\mathcal{G}} - \mathcal{G}\|_2^2),
\label{eq10}
\end{equation}
\begin{equation}
\mathcal{L}_{\text{diff}} = \mathbb{E}_{t \sim \mathcal{U}(1, T)} \mathcal{L}_t,
\label{eq:56}
\end{equation}
which regulates the predicted  $\widehat{\mathcal{G}}$ to approximate $\mathcal{G}$. In practical, we uniformly sample step $t$ at each training iteration to optimize $\mathcal{L}_{\text{diff}}$ over $t \sim \mathcal{U}(1, \ldots, T)$ (Eq. \ref{eq:56}). The procedure is detailed in Algorithm \ref{algorithmic3}. The loss function consists of the unified graph construction loss $\mathcal{L}_{\text{ugc}}$ and the diffusion loss $\mathcal{L}_{\text{diff}}$:
\begin{equation}
\mathcal{L} = \lambda_1\mathcal{L}_{\text{ugc}} + \mathcal{L}_{\text{diff}}, 
\label{eq:11}
\end{equation}
where the hyperparameter $\lambda_1$ serves as a weighting factor to balance the contributions of these two objectives.

\subsection{Inference}
\begin{table}[ht]
    \caption{Descriptive statistics of the datasets.}
    \label{tab:dataset}
    \centering 
    \resizebox{0.47\textwidth}{!}{
    \begin{tabular}{@{}lccc@{}} 
        \toprule
        & ML-1M & Yelp & Amazon-book \\ 
        \midrule
        \#Users & 5,949 & 54,574 & 108,822  \\ 
        \#Items & 2,810 & 34,395 & 94,949 \\ 
        \#Interaction & 571,531 & 1,402,736& 3,146,256  \\ 
        Density & 3.42\% & 0.07\% & 0.03\% \\ 
        \midrule
        Item Features & \begin{tabular}[c]{@{}l@{}}Title, Genres\end{tabular} & \begin{tabular}[c]{@{}l@{}}Stars, Text, \\ Useful, Cool\end{tabular} & \begin{tabular}[c]{@{}l@{}}Description, Price, \\ SalesRank, Categories\end{tabular} \\ 
        \bottomrule
    \end{tabular}}
\end{table}
\label{sectionE}
%
During the inference phase, the GDMCF takes the original bipartite graph $\mathcal{G}$ as input, which is corrupted by the addition of discrete and continuous noise, and outputs a clean bipartite graph. Specifically, GDMCF introduces noise into the original graph $\mathcal{G}$, resulting in the corrupted graphs $\mathcal{G}^{T}_R$ and $\mathcal{G}^{T}_X$. The corrupted graph $\mathcal{G}^{T}$, with aligned user features 
$\widebar{\mathbf{X}}^T$ and $\mathbf{R}^{T^\prime}$ generated through user-active guided generation, is then fed into a GCN-based denoiser for further generation. This network progressively denoises the graph, step by step, to predict the clean graph $\widehat{\mathcal{G}}$. Finally, the recovered graph $\widehat{\mathcal{G}}$ is utilized for item ranking, as detailed in Algorithm \ref{alg2}.


\section{Experiments}
\label{sec:experiment}
\subsection{Dataset Description}

We conduct experiments on three widely-used real-world datasets: 1) ML-1M \footnote{https://grouplens.org/datasets/movielens/1m/.}, a dataset containing movie ratings, 2) Yelp \footnote{https://www.yelp.com/dataset/.}, a service rating dataset where users share reviews and ratings and 3) Amazon-Book\footnote{https://jmcauley.ucsd.edu/data/amazon/.}, which contains reviews and book information from Amazon. For these datasets, we sort historical interactions by timestamp, remove users with fewer than four interactions, and split the datasets into training, validation, and test sets with a 7:1:2 ratio. Descriptive statistics for these datasets are shown in Tab. \ref{tab:dataset}.

\subsection{Experimental Setup}
\subsubsection{Evaluation metrics.} We use two evaluation metrics: 1) $Recall@K$ ($R@K$), measuring the proportion of relevant items in the top K recommendations. In Eq. \ref{eq:20}, $R(u)$ represents the set of top K recommendations for user $u$, while $T(u)$ is the set of items that user $u$ is interested in. 2) Normalized Discounted Cumulative Gain $NDCG@K$ ($N@K$), which considers the presence and position of relevant items. Eq. \ref{eq:21} shows the calculation, where $rel_i$ is the relevance score, $DCG$ is the ranking, and $iDCG$ is for normalization.
\begin{equation}
\begin{aligned}
Recall@K=\frac{\sum_{u \in \mathcal{U}}|R(u) \cap T(u)|}{\sum_{u \in \mathcal{U}}|T(u)|}
\end{aligned}
\label{eq:20}
\end{equation}

\begin{equation}
\begin{aligned}
NDCG@K=\frac{DCG@K}{iDCG}, \text { with }\left\{\begin{array}{l}
DCG@K=\sum_{i=1}^{K} \frac{rel_{i}-1}{\log _2(i+1)} \\
iDCG=\max _{\text {ranking }} DCG
\end{array}\right.
\end{aligned}
\label{eq:21}
\end{equation}

\subsubsection{Compared Methods.}
To demonstrate the effectiveness of GDMCF, we conducted a comparative analysis against nine methods, grouped as follows: 1) classical matrix factorization methods like MF \cite{rendle2009bpr};  2) GCN-based methods LightGCN \cite{he2020lightgcn}; 3) generative autoencoder methods, including MultiDAE \cite{liang2018variational}, CDAE \cite{wu2016collaborative}, and MultiVAE \cite{liang2018variational}; 4) diffusion generative methods like CODIGEM \cite{walker2022recommendation}, MultiDAE++ \cite{liang2018variational}, DiffRec \cite{wang2023diffusion} and CF-Diff \cite{DBLP:conf/sigir/HouPS24}.

\begin{itemize}
    \item \textbf{MF} \cite{rendle2009bpr} is a classical matrix factorization-based collaborative filtering method.
    \item \textbf{LightGCN} \cite{he2020lightgcn} is a lightweight graph convolutional network that generates node representations by aggregating information from neighboring nodes.
    \item \textbf{MultiDAE} \cite{liang2018variational} employs dropout to investigate a denoising autoencoder with a multinomial likelihood function.
    \item \textbf{CDAE} \cite{wu2016collaborative} trains an autoencoder to recover interactions that have been randomly corrupted.
    \item \textbf{MultiVAE} \cite{liang2018variational} employs variational autoencoders (VAEs) to recover interaction and uses Bayesian inference for parameter estimation.
     \item \textbf{CODIGEM} \cite{walker2022recommendation} models the diffusion process using multiple autoencoders (AEs), but only utilizes the first AE to predict interactions.
    \item \textbf{MultiDAE++} \cite{liang2018variational} incrementally adds Gaussian noise to interaction data and trains a MultiDAE to recover the interactions in a single step.
    \item \textbf{DiffRec} \cite{wang2023diffusion} applies Gaussian noise to each of the user's interactions and then reverses the process step by step.
    \item \textbf{CF-Diff} \cite{DBLP:conf/sigir/HouPS24}  is capable of making full use of collaborative signals along with multi-hop neighbors by a cross-attention-guided multi-hop autoencoder.
 \end{itemize}

\subsubsection{Hyper-parameter settings.}
We select the hyperparameters that yield the highest performance in terms of $NDCG@20$ on the test set. The learning rates are tuned among [0.0002, 0.001, 0.005, 0.01]. The batch size is fixed at 400, while the dimensionality of the latent embeddings and $\lambda_1$ are set to 1000 and 0.1, respectively. Additionally, we set the step embedding size of GDMCF to 10, the number of graph layers in GCN to 2, and the diffusion step $t$ within the range of [2, 5, 10, 20, 50, 100, 500]. The continuous noise scale is set within the range of 0.00001 to 0.25, and the discrete noise scale is set within the range of [0.0010, 0.0008, 0.0007, 0.0006, 0.0005, 0.0003]. Our experiments are conducted on a platform with an NVIDIA GeForce GTX 3090 GPU using PyTorch.

\label{sec:5.3}
\newcommand{\mrka}[1]{{\colorbox{red!30}{#1}}}  
\newcommand{\mrkb}[1]{{\colorbox{red!20}{#1}}}  
\newcommand{\mrkc}[1]{{\colorbox{red!10}{#1}}}  
\begin{table*}[t]
  \centering
  \caption{Performance comparison of the GDMCF framework and other baselines on three datasets. \% Improve. indicates the relative improvement of GDMCF over the best baseline results.}
  \resizebox{\linewidth}{!}{
    \begin{tabular}{lcccccccccccc}
      \toprule
      \multirow{2}{*}{Model} & \multicolumn{4}{c}{ML-1M} & \multicolumn{4}{c}{Yelp} & \multicolumn{4}{c}{Amazon-book}\\
      \cmidrule(lr){2-5} \cmidrule(lr){6-9} \cmidrule(lr){10-13}
       & $R@10 \uparrow$ & $R@20 \uparrow$ & $N@10 \uparrow$ & $N@20 \uparrow$ & $R@10 \uparrow$ & $R@20 \uparrow$ & $N@10 \uparrow$ & $N@20 \uparrow$ & $R@10 \uparrow$ & $R@20 \uparrow$ & $N@10 \uparrow$ & $N@20 \uparrow$ \\
      \midrule

    MF                        & 0.0876                                                     & 0.1503                                                     & 0.0749                                                     & 0.0966                                                     & 0.0341                                                     & 0.0560                                                     & 0.0210                                                     & 0.0276                                                     & 0.0437                                                     & 0.0689                                                     & 0.0264                                                     & 0.0339                                                     \\ 
LightGCN                  & 0.0987                                                     & 0.1707                                                     & 0.0833                                                     & 0.1083                                                     & 0.0540                                                     & 0.0904                                                     & 0.0325                                                     & 0.0436                                                     & 0.0534                                                     & 0.0822                                                     & 0.0325                                                     & 0.0411                                                     \\   
MultiDAE                  & 0.0995                                                     & 0.1753                                                     & 0.0803                                                     & 0.1067                                                     & 0.0522                                                     & 0.0864                                                     & 0.0316                                                     & 0.0419                                                     & 0.0571                                                     & 0.0855                                                     & 0.0357                                                     & 0.0442                                                     \\
CDAE                      & 0.0991                                                     & 0.1705                                                     & 0.0829                                                     & 0.1078                                                     & 0.0444                                                     & 0.0703                                                     & 0.0280                                                     & 0.0360                                                     & 0.0538                                                     & 0.0737                                                     & 0.0361                                                     & 0.0422                                                     \\ 
MultiVAE                  & 0.1007                                                     & 0.1726                                                     & 0.0825                                                     & 0.1076                                                     & 0.0567                                                     & 0.0945                                                     & 0.0344                                                     & 0.0458                                                     & \mrkc{0.0628}                                                     & \mrkc{0.0935}                                                    & \mrkc{0.0393}                                                     & \mrkc{0.0485}                                                     \\ 
CODIGEM                   & 0.0972                                                     & 0.1699                                                     & 0.0837                                                     & 0.1087                                                     & 0.0470                                                     & 0.0775                                                     & 0.0292                                                     & 0.0385                                                     & 0.0300                                                     & 0.0478                                                     & 0.0192                                                     & 0.0245                                                     \\
MultiDAE++                & 0.1009                                                     & 0.1771                                                     & 0.0815                                                     & 0.1079                                                     & 0.0544                                                     & 0.0909                                                     & 0.0328                                                     & 0.0438                                                     & 0.0580                                                     & 0.0864                                                     & 0.0363                                                     & 0.0448                                                     \\
DiffRec                   & \mrkc{0.1058 }                                                    & \mrkc{0.1787 }                                                    & \mrkc{0.0901 }                                                    & \mrkc{0.1148}                                               & \mrkc{0.0581}                                                     & \mrkc{0.0960}                                                     & \mrkc{0.0363}                                                     & \mrkc{0.0478}                                                     & \mrkb{0.0695}                                                     & \mrkb{0.1010}                                                     & \mrkb{0.0451}                                                     & \mrkb{0.0547}                                                     \\
    CF-Diff                   & \mrkb{0.1077}                                         & \mrkb{0.1843}                                         & \mrkb{0.0912}                                         & \mrkb{0.1176}                                         &  \mrkb{0.0585}                                         & \mrkb{0.0962}                                        & \mrkb{0.0368}                                         & \mrkb{0.0480}                                         &  0.0499                                                         & 0.0717                                                         & 0.0337                                                         & 0.0404                                                         \\ 
      GDMCF                    & \mrka{0.1078}                                            & \mrka{0.1861}                                            & \mrka{0.0916}                                            & \mrka{0.1178}                                            & \mrka{0.0634}                                            & \mrka{0.1044}                                            & \mrka{0.0392}                                            & \mrka{0.0515}                                            & \mrka{0.0916}                                            & \mrka{0.1315}                                            & \mrka{0.0587}                                            & \mrka{0.0707}                                            \\
       \hline
    \% Improve.               & 0.09\%                                                     & 0.98\%                                                     & 0.44\%                                                     & 0.17\%                                                     & 8.38\%                                                    & 8.52\%                                                    & 6.52\%                                                    & 7.29\%                                                    & 31.80\%                                                          & 30.20\%                                                          & 30.16\%                                                          & 29.25\%                                                          \\ 
      \bottomrule
     
    \end{tabular}
  }
  
  \label{performance}
\end{table*}

\subsection{Performance Comparisons}
We summarized the performance of various methods on three datasets in terms of $Recall@K$ and $NDCG@K$ ($K=10,20$). As shown in Tab. \ref{performance}, GDMCF outperforms all the baselines across all metrics. Additionally, we have the following observations:

\begin{itemize}
    \item The matrix factorization method (MF) demonstrates the weakest performance among all baselines. This is primarily because MF relies on explicit user-item interactions, overlooks implicit user preference information, and fails to capture higher-order relationships. 
    \item Most generative models outperform discriminative models (MF and LightGCN) due to their superior capacity to model the joint distribution of complex latent factors. 
    
    \item Among generative baselines, CF-Diff and DiffRec outperform others, highlighting the effectiveness of the diffusion process and the step-by-step denoising approach for recommendation tasks. CODIGEM performs worse, likely due to its sole reliance on the first AE for inference.
   
    \item GDMCF achieves state-of-the-art performance on all datasets by effectively capturing higher-order collaborative signals between users and items during denoising, enabling better generative recommendations.
    
    \item User interaction data in real-world recommender systems often provides only the final click results, while the underlying decision-making process remains highly complex and influenced by a diverse set of factors. GDMCF models the heterogeneous noise present in real-world recommendation scenarios by explicitly capturing the inherent noise embedded within user interaction data and improves the overall recommendation performance.
    
    \item On large-scale datasets, GDMCF demonstrates a more pronounced advantage. This is because traditional generative methods primarily focus on element-wise user interaction generation, making it difficult to capture the complex interdependencies among nodes in large bipartite graphs. In contrast, GDMCF works at the full-graph level, making it better suited for large-scale scenarios.   
\end{itemize}


\subsection{Ablation Studies}

\begin{table}[]
\caption{Influence of Multi-level Corruption on Yelp.}
\label{Ablation1}
\resizebox{0.9\linewidth}{!}{
\begin{tabular}{ccccc}
\toprule
Model & $R@10$   & $R@20$   & $N@10$   & $N@20$   \\ \hline
GDMCF$_{CC}$    & 0.0571 & 0.0926 & 0.0355 & 0.0462 \\
GDMCF$_{DC}$   & 0.0578 & 0.0952 & 0.0351 & 0.0464 \\
GDMCF$_{NC}$  & 0.0548 & 0.0893 & 0.0333 & 0.0438 \\ \hline
GDMCF & 0.0634 & 0.1044 & 0.0392 & 0.0515 \\ \bottomrule
\end{tabular}}
\end{table}
\begin{table}[]

\caption{Influence of User-active Guided Generation strategy.}
\label{Ablation2}
\resizebox{0.9\linewidth}{!}{
\begin{tabular}{cclcl}
\toprule
Model                  & $R@10$                                & \multicolumn{1}{c}{$R@20$} & $N@10$                               & \multicolumn{1}{c}{$N@20$} \\ \hline
GDMCF$_{NUA}$                     & \multicolumn{1}{l}{0.0259}          & 0.0420          & \multicolumn{1}{l}{0.0205}         & 0.0261                   \\
GDMCF                  & \multicolumn{1}{l}{\textbf{0.0267}} & \textbf{0.0427}                   & \multicolumn{1}{l}{\textbf{0.0208}} & \textbf{0.0264}          \\ \hline
\multirow{2}{*}{Model} & \multicolumn{4}{c}{Inference}                                                                                                  \\ \cline{2-5} 
                       & \multicolumn{2}{c}{Speed (s)}                                  & \multicolumn{2}{c}{Memory (MiB)}                              \\ \hline
GDMCF$_{NUA}$                     & \multicolumn{2}{c}{204}                                        & \multicolumn{2}{c}{12256}                                     \\ \hline
GDMCF                  & \multicolumn{2}{c}{\textbf{159}}                                        & \multicolumn{2}{c}{\textbf{7486}}                                     \\ 
\bottomrule
\end{tabular}}
\end{table}

\begin{figure}[htbp]
    \centering
    \begin{subfigure}{0.23\textwidth}
        \centering
        \includegraphics[width=\linewidth]{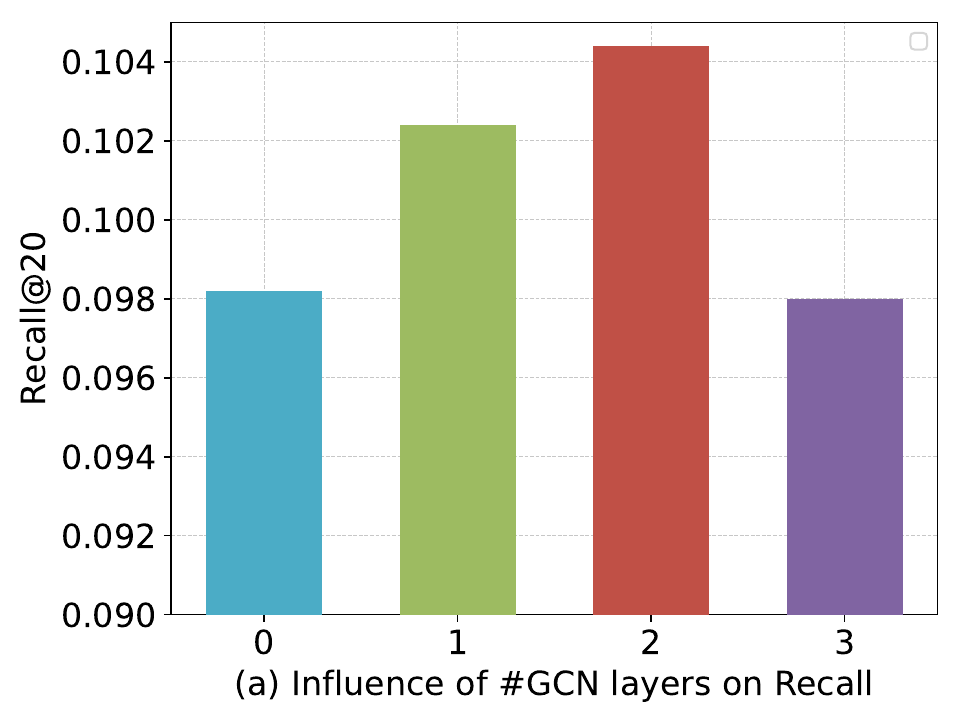}
        \label{fig:sub1}
    \end{subfigure}
    \hfill
    \begin{subfigure}{0.23\textwidth}
        \centering
        \includegraphics[width=\linewidth]{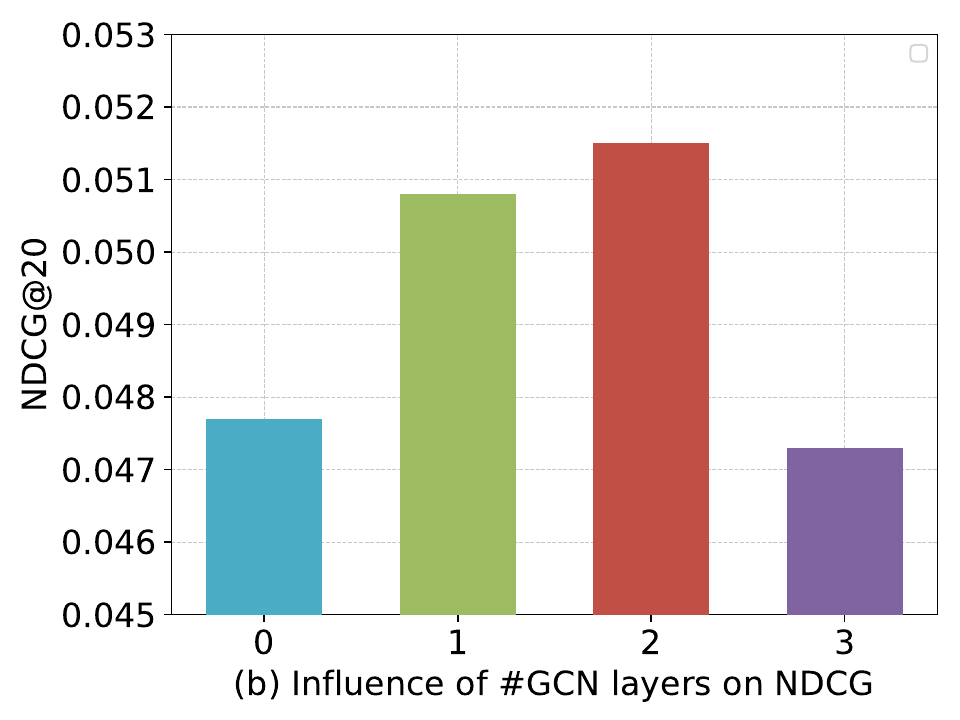}
        \label{fig:sub2}
    \end{subfigure}
    \caption{Influence of the number of  GCN layers $l$.}
    \label{fig:main_figure2}
\end{figure}

\begin{figure}[htbp]
    \centering
    \begin{subfigure}{0.23\textwidth}
        \centering
        \includegraphics[width=\linewidth]{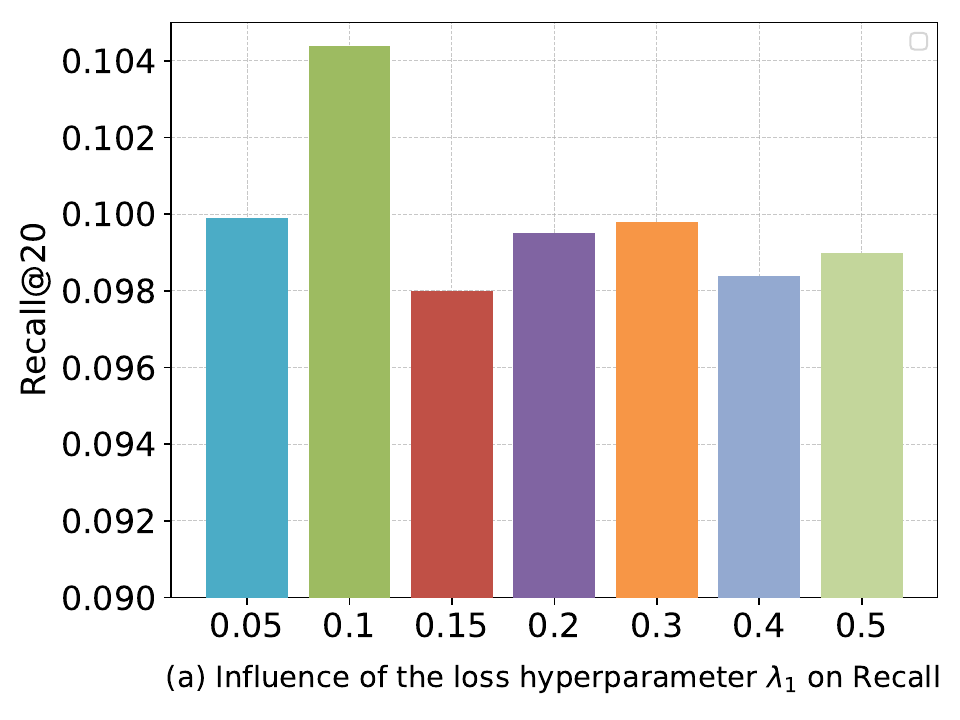}
        \label{fig:sub1}
    \end{subfigure}
    \hfill
    \begin{subfigure}{0.23\textwidth}
        \centering
        \includegraphics[width=\linewidth]{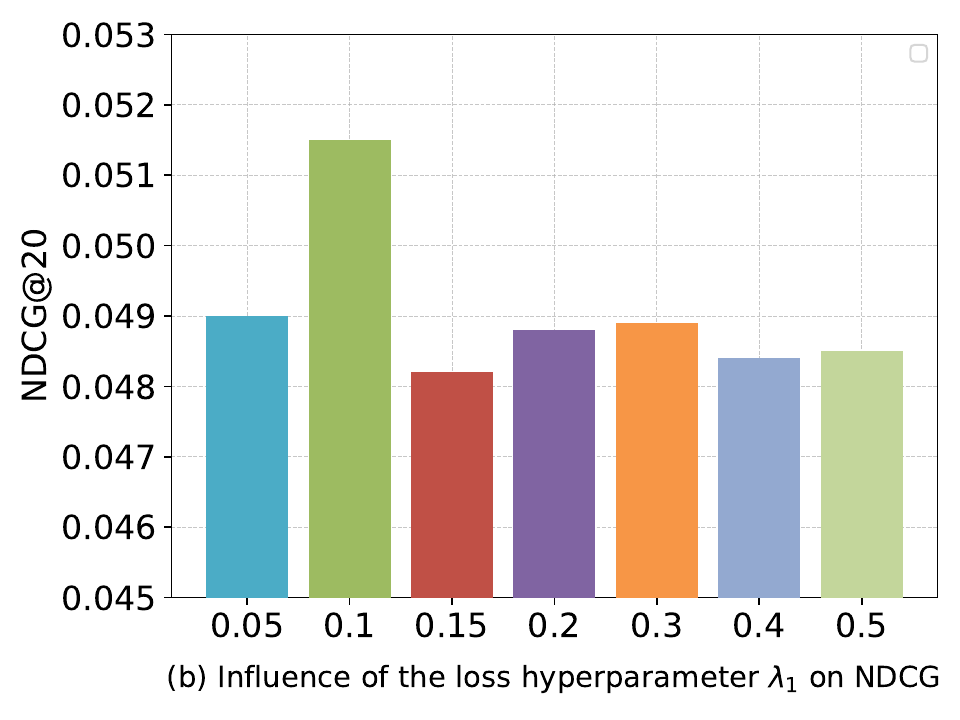}
        \label{fig:sub2}
    \end{subfigure}
    \caption{Influence of the loss hyperparameter $\lambda_1$.}
    \label{fig:sub3}
\end{figure}

\subsubsection{Influence of Multi-level Corruption.}
\begin{figure}[htbp]
    \centering
    \begin{subfigure}{0.23\textwidth}
        \centering
        \includegraphics[width=\linewidth]{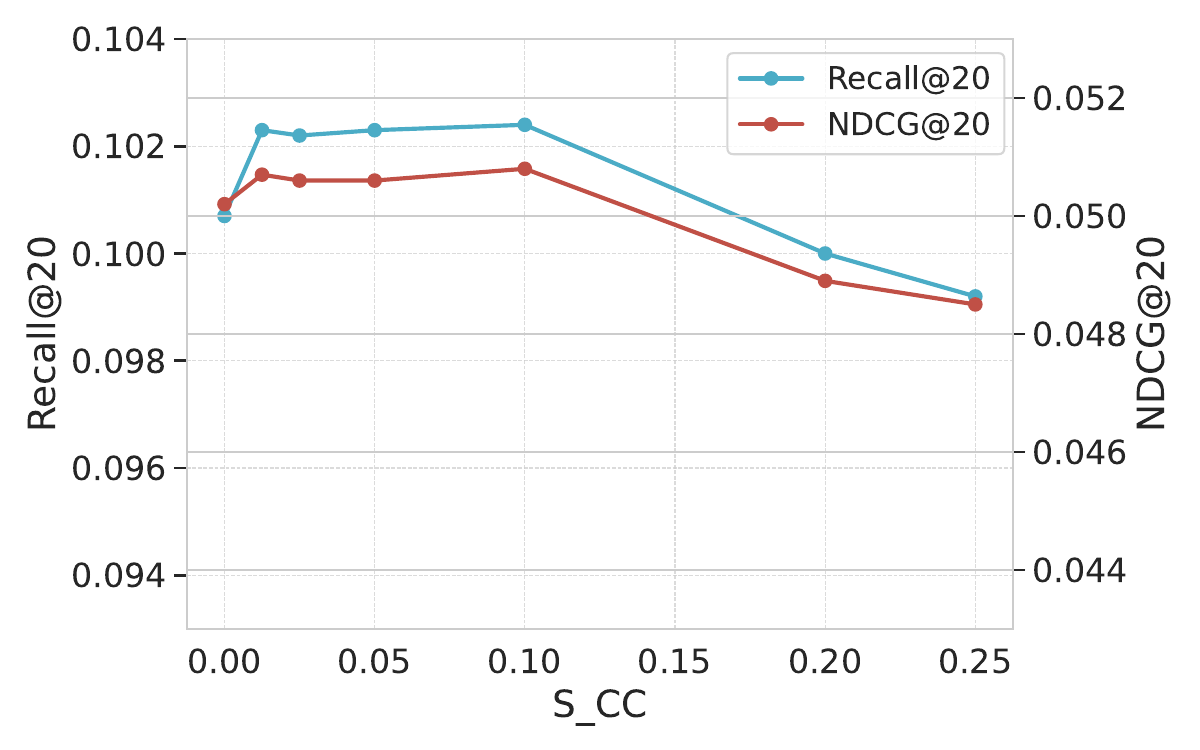}
        \label{fig:sub34}
    \end{subfigure}
    \hfill
    \begin{subfigure}{0.23\textwidth}
        \centering
        \includegraphics[width=\linewidth]{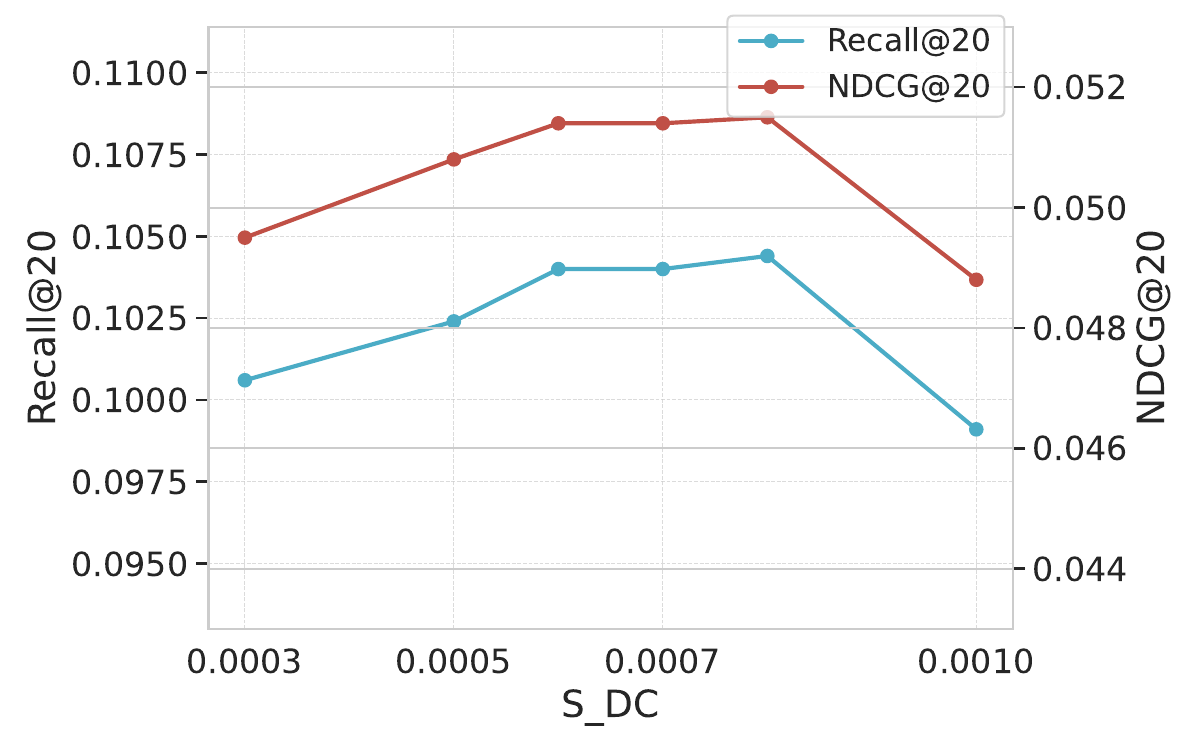}
        \label{fig:sub4}
    \end{subfigure}
   \vspace{-3mm} 
    \caption{Influence of the noise scale $S_{CC}$ and $S_{DC}$.}
    \label{fig:main_figure}
\end{figure}
To investigate the influence of multi-level corruption, we considered several variants of GDMCF:
\begin{itemize}
 \item GDMCF$_{CC}$: In the Multi-level Corruption stage, we retained only continuous corruption.
 \item GDMCF$_{DC}$: In the Multi-level Corruption stage, we retained only discrete corruption.
 \item GDMCF$_{NC}$: We removed the diffusion process, directly inputting the original topology graph into the GCNs.
\end{itemize}
Tab. \ref{Ablation1} shows the impact of different corruptions on Yelp. Removing either discrete corruption (GDMCF$_{CC}$) or continuous corruption (GDMCF$_{DC}$) leads to a decline in performance and even causes training instability. This is likely because diverse noise in recommendation scenarios requires multiple diffusion levels to fully capture it, and using just one makes the model less accurate. Without the diffusion process, GDMCF reduces to a model that inputs the original graph directly into the GCNs, omitting the critical step-by-step denoising process (GDMCF$_{NC}$). GDMCF operates as a Markov process with graph edits, is permutation equivariance, and provides an evidence lower bound for likelihood estimation. These properties likely explain its superior performance compared to GDMCF$_{NC}$.


\subsubsection{Efficiency of User-active Guided Generation.}


We sampled 3,000 users and their associated items, creating a subdataset from the Yelp dataset, referred to as Yelp-s. This subgraph was utilized to assess the impact of GDMCF and its variant, GDMCF$_{NUA}$, on computational memory efficiency.  In GDMCF$_{NUA}$, the user-active guided  strategy is canceled, 
meaning that during inference, the absence of user-active nodes diminishes the differentiation in the significance of predicted edges. These redundant edges not only result in a loss of topological information but also introduce additional computational overhead. This could explain why GDMCF outperforms GDMCF$_{NUA}$, as demonstrated in Tab. \ref{Ablation2}.

\subsection{Parameter Sensitivity Analysis}
We analyze GDMCF's hyperparameters, focusing on noise scale, GCN layers, and loss factor $\lambda_1$. Due to space constraints, we present only the $R@20$ and $N@20$ metrics from the Yelp dataset.


\subsubsection{GCN layers}
We investigated the influence of the number of GCN layers, denoted as \( l \), on performance. When \( l = 0 \), the GCN reduces to a multilayer perceptron (MLP), which is inadequate for processing complex graph information, resulting in suboptimal performance. As the number of GCN layers increases, performance initially improves but subsequently declines, indicating that excessive higher-order information may introduce unnecessary noise. Detailed results are presented in Fig. \ref{fig:main_figure2}.

\subsubsection{Sensitivity Analysis of $\lambda_1$}
To explore the sensitivity of GDMCF to the hyperparameter $\lambda_1$, we conducted a series of experiments varying $\lambda_1$. The results indicate that the recommendation performance is optimal when $\lambda_1$ = 0.1. Furthermore, as $\lambda_1$ increases, neither Recall nor NDCG exhibits any significant improvement. Detailed results are presented in Fig. \ref{fig:sub3}.

\subsubsection{Noise Scale} 
$S_{CC}$ and $S_{DC}$ are the noise scales for continuous and discrete corruption, respectively. As shown in Fig. \ref{fig:main_figure}, GDMCF achieves optimal performance at $S_{CC}=0.1$ and $S_{DC}=0.0008$. However, performance declines with increasing \( S_{CC} \). An appropriate \( S_{DC} \) must align with the original graph's topology, as excessively large or small values can disrupt the contained information.
\begin{figure}[htbp]
    \centering
    \begin{subfigure}{0.224\textwidth}
        \centering
        \includegraphics[width=\linewidth]{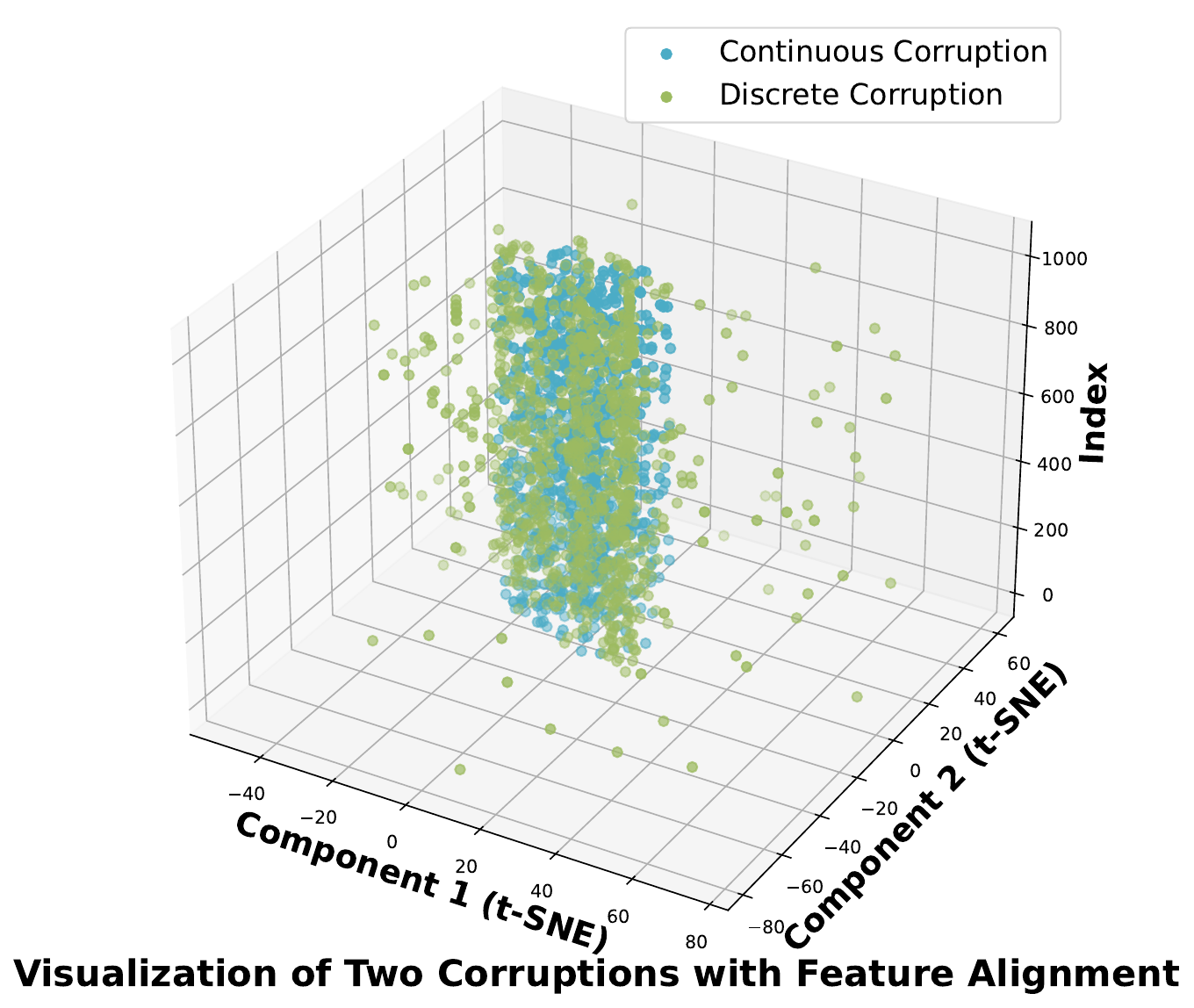}
        \label{fig:sub1}
    \end{subfigure}
    \hfill
    \begin{subfigure}{0.236\textwidth}
        \centering
        \includegraphics[width=\linewidth]{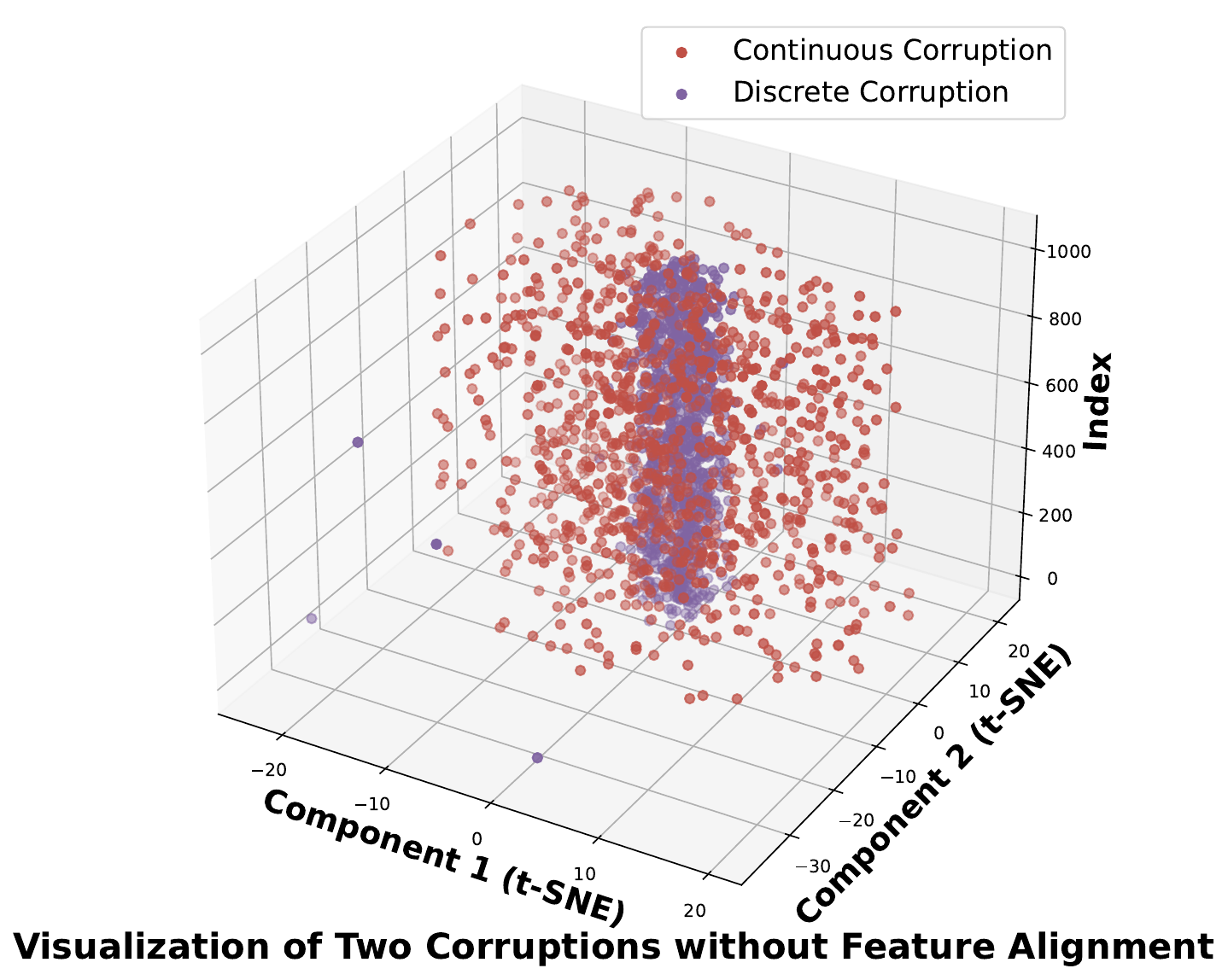}
        \label{fig:sub2}
    \end{subfigure}
   \vspace{-3mm}
    \caption{Feature alignment visualizition of two corruptions.}
    \label{fig:visual}
\end{figure}
\subsection{Additional Analysis}
\begin{table}[ht]
\centering
\caption{Evaluation on long-tail distribution.}
\label{Ablation4}
\begin{tabular}{lcccc}
\toprule
\textbf{Methods} & \textbf{R@10} & \textbf{R@20} & \textbf{N@10} & \textbf{N@20} \\ 
\midrule
Average          & 0.1050             & 0.1795             & 0.0867           & 0.1127           \\
GDMCF            & \textbf{0.1120}    & \textbf{0.1797}    & \textbf{0.0920}  & \textbf{0.1164}  \\ 
\bottomrule
\end{tabular}
\end{table}
\subsubsection{Evaluation on long-tail distribution}

In user-active guided generation, GDMCF incrementally adds edges to an initially empty graph based on probability, ensuring a diverse graph structure. Unlike fixed graph architectures where highly popular users tend to dominate, this approach provides less popular (long-tail) users with a probability of being activated, thereby increasing their exposure. This probabilistic sampling enhances the model's robustness and helps alleviate potential popularity bias. We conducted experiments on the Yelp dataset, focusing on the long-tail distribution. Specifically, we identified the bottom 20\% of users based on their interaction frequency and compared the performance of our user-active guided strategy against that of a baseline employing an average probability-based strategy for this subgroup of users. The results in Tab. \ref{Ablation4} illustrate that GDMCF effectively reduces popularity bias and offers better recommendations for long-tail users.

\subsubsection{Feature alignment visualizition}
To verify the efficiency of the unified graph construction module that aligns the feature-level and topological-level attributes within the bipartite graph, we visualized the unaligned features$\boldsymbol{X}^{\prime} $ and \( \boldsymbol{X}^{\prime \prime} \), as well as the aligned features $\boldsymbol{X}^{\prime} $ and \( \boldsymbol{X}^{\prime \prime} \) on  Yelp. Fig. \ref{fig:visual} demonstrates that, after alignment, the features of the aligned users became significantly closer.

\section{Conclusion}

\label{sec:conclusion}

Generative models outperform discriminative models by capturing the joint distribution of complex latent factors, and diffusion-based recommendation methods have shown remarkable results. However, existing approaches operate primarily at the element-wise level, overlooking higher-order collaborative signals. To address this, we propose GDMCF, which captures higher-order collaborative signals at the graph level, thereby improving graph-based diffusion learning. To tackle noise heterogeneity, we employ multi-level corruption and align the corrupted features into a unified space for graph-based denoising. Additionally, we reduce inference costs by retaining only the edges associated with activated users in the bipartite graph. Extensive experiments on three datasets demonstrate that GDMCF consistently outperforms competing methods in both effectiveness and efficiency.



\begin{acks}
This research has been partially supported by the National Nature Science Foundation of China (No.62173199).
\end{acks}

\bibliographystyle{ACM-Reference-Format}
\bibliography{sample-main}


\end{document}